\documentclass[10pt,draftcls,onecolumn,fleqn]{IEEEtran}
\IEEEoverridecommandlockouts  

\usepackage{amssymb}
\usepackage{amsmath}
\usepackage{amsfonts}                                
\usepackage[latin1] {inputenc}
\usepackage{enumerate}
\usepackage{mathrsfs}
\usepackage{tabulary}
\usepackage{cite}
\usepackage{psfrag}
\usepackage{graphicx}          
\usepackage{color}      
\usepackage{epsfig} 
\usepackage{times} 
\usepackage{bigints}

\def\qed{\relax\ifmmode\hskip2em \Box\else\unskip\nobreak\hskip1em $\Box$\fi}

\newtheorem{theorem}{Theorem}
\newtheorem{itlemma}{Lemma}
\newtheorem{itdefinition}{Definition}
\newtheorem{itproposition}{Proposition}
\newtheorem{itresult}{Result}
\newtheorem{itremark}{Remark}
\newtheorem{itassumption}{Assumption}
\newtheorem{itcorollary}{Corollary}
\newtheorem{itexample}{Example}

\newenvironment{proposition}{\begin{itproposition}\rm}{\end{itproposition}}

\def\qed{\hfill\vrule height 1.6ex width 1.5ex depth -.1ex}
\newcommand{\be}{\begin{equation}}
\newcommand{\ee}{\end{equation}}
\newcommand{\ben}{\begin{equation*}}
\newcommand{\een}{\end{equation*}}
\newcommand{\ba}{\begin{array}}
\newcommand{\ea}{\end{array}}

\newcommand{\defi} { \stackrel{\bigtriangleup}{=} }

\title{\textbf{Event-triggered distributed Bayes filter}} 

\author{Giorgio Battistelli, Luigi Chisci, Lin Gao, and Daniela Selvi
\thanks{
G. Battistelli, L. Chisci, and L. Gao are  with Dipartimento di Ingegneria dell'Informazione (DINFO), Universit$\grave{\mbox{a}}$ degli Studi di Firenze, Via Santa Marta 3, 50139, Firenze, Italy (Email: giorgio.battistelli@unifi.it; luigi.chisci@unifi.it; lin.gao@unifi.it).
D.Selvi is with Dipartimento di Ingegneria Industriale (DIEF), Universit$\grave{\mbox{a}}$ degli Studi di Firenze, Via Santa Marta 3, 50139, Firenze, Italy (Email: daniela.selvi@unifi.it).
}
}

\begin{document}

\maketitle           

\begin{abstract}
	The aim of this paper is to devise a strategy that
	is able to reduce communication bandwidth and, consequently, energy consumption 
	in the context of distributed state estimation over a peer-to-peer sensor network. 
	Specifically,
	a distributed Bayes filter with event-triggered communication is developed 
	by enforcing each node to transmit its local information to the neighbors 
	only when the Kullback-Leibler divergence between the current local posterior 
	and the one predictable from the last transmission exceeds a preset threshold. 
	The stability of the proposed event-triggered distributed Bayes filter is proved in the linear-Gaussian (Kalman filter) case.
	The performance of the proposed algorithm is also evaluated through simulation experiments concerning a target tracking application.
\end{abstract}

\section{Introduction} \label{sec:introduction}
The problem of \textit{distributed state estimation} (DSE) on a \textit{wireless sensor network} (WSN) has attracted considerable attention
due to its wide and successful applicability to many distributed monitoring tasks in the industrial, environmental and defense contexts \cite{dargie2010fundamentals}.
In this respect, several approaches to DSE have been developed such as, for instance, the distributed Kalman filter (KF) \cite{olfati2009kalman,das2015distributed,das2017consensus} for the linear case  or
the distributed extended KF \cite{battistelli2016stability,battistelli2015consensus}, distributed unscented KF \cite{li2016weighted} and distributed particle filter (PF) \cite{hlinka2012likelihood} for the nonlinear case.

Normally, sensor nodes of WSNs are battery-powered and, thus,
have limited energy.
Hence it is of paramount importance to reduce the message transmission between sensor nodes (i.e., the communication rate)
in order to save energy.
Another motivation for reducing message transmission is in defense applications,
where each message transmission increases the risk of discovery of sensor nodes.
Generally speaking,
the reduction of message transmission 
can be accomplished by resorting to an \textit{event-triggered} (ET) strategy \cite{battistelli2018distributed}, by which a suitable triggering test is carried out at each sensor node to check in advance whether it is worth transmitting a given message or not.

In centralised multisensor systems, 
ET strategies have been successfully exploited to reduce the communication bandwidth \cite{han2015stochastic,marck2010relevant,trimpe2014event,battistelli2012data}.
Recently, attempts have also been carried out to apply ET strategies in distributed state estimation with satisfactory results. 
In \cite{li2016event}, the information is transmitted by each sensor node
whenever the distance between the most recently transmitted estimate 
and the current one exceeds a pre-defined threshold, 
where the distance is measured in terms of the mean square error (MSE),
while the second-order moment (covariance) discrepancy is ignored. 
In \cite{yan2014distributed},
each sensor node broadcasts a local measurement to the neighbors 
only when its \textit{Mahalanobis distance} (MD) from the latest transmitted measurement exceeds a given threshold.
However,  as shown in \cite{battistelli2015consensus} and \cite{battistelli2014kullback}, such DSE algorithms exchanging measurements among sensor nodes
 cannot guarantee stability unless the number of data exchanges is large enough.
In our recent work \cite{battistelli2018distributed}, 
an ET strategy is proposed along with a consensus method for DSE
 with guaranteed stability. 
At each sensor node,  transmission of local information to the neighbors is triggered whenever the local estimate and/or covariance deviate from the ones predicted after the
last transmission of a sufficiently high amount.

In this paper, the aim is to develop an ET-DSE approach following a Bayesian filtering perspective for DSE
\cite{battistelli2014parallel} 
and adopting an information-theoretic criterion for transmission triggering.
In particular,
it is assumed that each node, besides the local \textit{probability density function} (PDF), stores the last transmitted (reference) PDF and also the last received (neighbor) PDFs from all neighbors.
Then, after each local update and before consensus, message sending is triggered whenever
the \textit{Kullback-Leibler Divergence} (KLD), aka information gain, from the predicted reference PDF to the
local posterior PDF exceeds a given threshold.
In a consensus step,
when a node does not receive a message from some neighbor,
it can recover the local posterior PDF of such a neighbor
with satisfactory accuracy via prediction of the stored neighbor PDF.
The rationale of this recovery is that,
if a neighbor does not transmit messages,
its local posterior must be sufficiently close to the predicted reference PDF.
The advantage of the proposed ET \textit{consensus Bayes filter} (ET-CBF)  is that 
the communication bandwidth/energy consumption of each sensor node can be significantly reduced 
while deteriorating the tracking performance as least as possible. 
Compared to the standard CBF, 
the proposed ET-CBF just needs little extra memory space for storing reference 
as well as neighbors' information. 
Moreover,
the proposed ET-CBF can also be regarded as a generalization of the ET-KF presented in \cite{battistelli2018distributed}.

Similar ET strategies have also also been proposed to handle 
the problem of distributed joint detection and tracking of a target,
which results in the so-called ET consensus Bernoulli filter \cite{gao2018event}.
            
The rest of the paper is organized as follows.
Section II reviews consensus-based distributed Bayesian filtering.
Section III introduces the KLD-based ET criterion and develops the proposed ET-DSE algorihm.
Section IV analyses its stability in the linear-Gaussian case.
Section V provides a performance evaluation of the proposed ET-DSE via simulation experiments concerning a target tracking case study.
Finally, section VI ends the paper with some concluding remarks.

\section{Distributed state estimation with consensus on posteriors} 

This paper addresses DSE over a network in which 
each node can process local data as well as exchange data with neighbors. 
Further, some nodes can also sense data from the environment, and are called sensor nodes. The task of nodes without sensing capabilities, called communication nodes, is only to improve network connectivity.
In the sequel, the sensor network will be denoted as $\left( \mathcal{N}, \mathcal{A}, \mathcal{S} \right)$ where:
$\mathcal{N} = \{1, \ldots, N \}$ is the set of nodes;
$\mathcal{A} \subseteq \mathcal{N} \times \mathcal N$ is the set of arcs (edges) such that $(i,j) \in \mathcal{A}$ if node $j$ can receive data from node $i$;
$\mathcal S \subseteq \mathcal N$ is the subset of sensor nodes. 
Further, for each node $i \in \mathcal{N}$, $\mathcal{N}_i \subseteq \mathcal N$ will denote the set of its in-neighbors, i.e.
$\mathcal{N}_i \defi \left\{ j: (j,i) \in \mathcal{A} \right\}$.

The DSE problem can be formulated as follows.
Each node $i \in \mathcal{N}$ must estimate at each time $k \in \{ 0, 1, \dots \}$ the state $x_k$ of the dynamical system
\begin{eqnarray}
x_{k+1} = f_k (x_k) + w_k \label{NLS1}
\end{eqnarray}
given local measurements 
\begin{eqnarray}
y^i_{k} & = & h_k^i (x_k) + v^i_k \, , ~~~ i \in \mathcal{S} \, ,
\label{NLS2}
\end{eqnarray}
and data received from all neighboring nodes $j \in \mathcal{N}_i$.
The initial state $x_0$ and the sequences $\{ w_k \}$ and $\{ v_k^i \}$, representing respectively the process disturbance and measurement noises, are supposed to be mutually independent.
The sequence $\{ w_k \}$ is supposed to be generated by a white stochastic process
with known PDF $p_w(\cdot)$. Similarly, $\{ v^i_k \}$ is generated by a white stochastic process with known PDF $p_{v^i} (\cdot)$.

Consider first the case in which no information exchange is performed among the network nodes, i.e., each node independently runs its own local filter so as to estimate the state $x_k$.
As well known, in this case, the solution of the local state estimation problem
would yield the Bayes filter recursion:
\begin{eqnarray}
p^i_{k|k} (x) &=& \frac{p_{v^i} (y_k^i - h^i_k (x)) \, p^i_{k|k-1} (x)}{ \int p_{v^i} (y_k^i - h^i_k (\xi)) \, p^i_{k|k-1} (\xi) d \xi} \, ,  \label{eq:bayes:1} \\
p^i_{k+1|k} (x) &=& \int p_{w} (x - f_k(\xi))  \, p^i_{k|k} (\xi) d \xi \, , \label{eq:bayes:2}
\end{eqnarray}
for $k=0,1,\ldots$,
where $p^i_{k|t}(\cdot)$ represents the PDF of $x_k$ conditioned to all the measurements collected by node $i$ up to time $t$, and the
recursion is initialized at time $k=0$ from some prior density $p^i_{0|-1} (x) $.

Suppose now that  a communication structure is available as described previously so that 
each node $i$ can receive data from the nodes belonging to the subset $\mathcal N_i \subseteq \mathcal N$.
Then, in order to improve its local estimate, each node $i$ can fuse the local information, i.e., the local posterior $p^i_{k|k} (\cdot)$,
with the one received from its neighbors  $p^j_{k|k} (\cdot)$, $j \in \mathcal N_i$. More specifically, 
one can perform at each time instant a certain number, say $L$, of consensus steps on the posterior PDFs
$p^i_{k|k} (\cdot) , \, i \in \mathcal N$, in order to compute in a distributed fashion their average. 
This can be done by following the approach of \cite{battistelli2014kullback}.
More specifically, consider a generic node $i$ at time $k$ 
and suppose that $\ell$ consensus iterations have been carried out yielding
the posterior density $p_{k,\ell }^i(x)$.
Then, the fused density at the next consensus step $p_{k,\ell +1 }^i(x)$ is obtained by computing a normalized geometric mean among the local density and those of the neighbors
\begin{eqnarray}
p_{k,\ell  + 1}^i\left( x \right) = \frac{{  \left [ p^i_{k,\ell} (x) \right ]^{\pi_{i,i}} \displaystyle{\prod\limits_{j \in {{\cal N}_i}} {{{\left[ {p_{k,\ell }^j\left( x \right)} \right]}^{{\pi_{i,j}}}}} }}}{\displaystyle{\int { \left [ p^i_{k,\ell} (x) \right ]^{\pi_{i,i}} \prod\limits_{j \in {{\cal N}_i}} {{{\left[ {p_{k,\ell }^j\left( x \right)} \right]}^{{\pi_{i,j}}}}} dx} }}   
 \label{eq:GCI}  
\end{eqnarray}
where the consensus weights $\pi_{i,j}$ must satisfy ${\pi_{i,j}} > 0$ and $\pi_{i,i} + \sum\nolimits_{j \in {{\cal N}_i}} {{\pi_{i,j}}}  = 1$.
Clearly, in each network node $i$ the consensus recursion is initialized from the local posterior densities 
by setting $p_{k,0} (x) = p^i_{k|k} (x)$. As discussed in \cite{battistelli2014kullback}, the fusion rule (\ref{eq:GCI}) (known in the literature as {\em Generalized Covariance Intersection}) has a meaningful interpretation as the average, 
in terms of {\em Kullback-Leibler Divergence},  of the densities to be fused. For this reason, it has also been referred to as {\em Kullback-Leibler average}.
An important property of the consensus algorithm based on the fusion rule  (\ref{eq:GCI})  is that, under suitable assumptions \cite{battistelli2014kullback}, as the number $\ell$ of consensus steps increases all the local densities 
converge to the collective average
\begin{eqnarray}
p_{k} \left( x \right) = \frac{\displaystyle{\prod\limits_{j \in {{\cal N}}} {{{\left[ {p_{k|k }^j\left( x \right)} \right]}^{1/N}}} }}{\displaystyle{\int {\prod\limits_{j \in {{\cal N}}} {{{\left[ {p_{k|k}^j\left( x \right)} \right]}^{1/N}}} dx} }}   
 \label{eq:KLA}  
\end{eqnarray}
Summing up, the DSE algorithm of Table \ref{tab:1} is obtained.

\begin{table}[tb]
\caption{Algorithm 1 - Distributed State Estimation with Consensus on Posteriors} \label{tab:1}
\hrulefill \\
At each time $k=0,1,\ldots$, for each node $i \in \mathcal N$:
\begin{enumerate}
\item {\bf Correction:} \\ {\bf If} $i \in \mathcal S$, collect the local measurement $y_k^i$ and update the local prior
$p^i_{k|k-1} $ via equation (\ref{eq:bayes:1})
to obtain the local posterior $p^{i}_{k|k}$; \\
otherwise, for any $i \in \mathcal N \setminus \mathcal S$, set
$p^{i}_{k|k} = p^i_{k|k-1} $;
\item {\bf Consensus:} \\ 			
set $p^i_{k,0} = p^i_{k|k} $; \\
 {\bf For} $\ell  = 0, \ldots ,L-1$,  \\
		\mbox{}	\quad \quad transmit $p_{k,\ell}^i$ to the out-neighbors; \\			
		\mbox{}	\quad \quad receive  $p_{k,\ell}^j$  from the in-neighbors $j \in {{\cal N}_i}$; \\	
		\mbox{}	\quad \quad	perform fusion using (\ref{eq:GCI}); \\
		{\bf End for}
\item {\bf Prediction:} \\ compute the local prior $p^i_{k+1|k}$ from $p^{i}_{k,L}$ via equation (\ref{eq:bayes:2}).
\end{enumerate}
\hrulefill
\end{table}

Notice that in principle Algorithm 1 can deal with PDFs of arbitrary form. Clearly, a closed-form
expression for the recursion exists only in special cases (for instance, when the system dynamics and measurement equations are linear and all the random variables are Gaussian).
Hence, in general, the treatment of a nonlinear and/or non-Gaussian
setting requires some sort of approximation, for example based on  the Extended Kalman Filter (EKF) \cite{battistelli2016stability} or the Unscented Kalman filter (UKF) \cite{battistelli2014parallel}.
Algorithm 1 can be modified, by introducing suitable correcting factors, in order to weight differently the prior and novel information in the information fusion step, so as to reduce conservativeness 
while preserving stability \cite{IWC,battistelli2015consensus,battistelli2014parallel}.  

As a final remark, it is worth pointing out that Algorithm 1 and its variants enjoy nice stability properties. 
In fact, in \cite{battistelli2014kullback,battistelli2015consensus} it has been shown that, irrespectively of the number $L$ of consensus steps, 
Algorithm 1 ensures a mean-square-bounded estimation error in each network node provided that  the system is {\em collectively observable} and the network is {\em strongly connected}.
The stability result can also be extended to the nonlinear case when the Extended Kalman filter is used in the correction/prediction steps \cite{battistelli2016stability}.

\section{Event-triggered distributed Bayes filter}\label{sec.problem}

In Algorithm 1, it is supposed that, at every discrete time instant $k$, each node $i \in \mathcal N$ sends the local density to its out-neighbors (even multiple times when $L>1$).
However, in many contexts, it is desirable to reduce data transmission as much as possible while preserving stability and performance. This goal can be achieved by controlling transmission so that
each node $i$ selectively transmits only the most relevant data.
To this end, let us introduce for each node $i$ binary variables $c^i_{k,\ell}$
such that $c^i_{k,\ell} = 1$ if node $i$ transmits at time $k$ and consensus step $\ell$, or $c^i_{k,\ell} = 0$ otherwise. 
We focus on data-driven transmission strategies in which the variable $c^i_{k,\ell}$
is a function of $p^i_{k,\ell}$ (the local density currently available in node $i$) and of the density most recently transmitted by node $i$.

Let us now denote by $\bar p^i_{k,\ell}$ the so-called {\em reference density}, obtained by propagating the most recently transmitted density up to the current time instant.
Clearly, this means that, in case the last transmission has occurred at time $k' < k$, the reference density is obtained from the most recently transmitted density via $k-k'$ prediction steps. Conversely,
if the last transmission has occurred at time $k$, the reference density simply coincides with the most recently transmitted one.
Noting that the reference density $\bar p^i_{k,\ell}$ can be computed also by the out-neighbors of node $i$,
the idea is that, when  the discrepancy between $p^i_{k,\ell}$ and $\bar p^i_{k,\ell}$ is small, we do not really need to transmit  $p^i_{k,\ell}$ because the information gain
obtained by replacing $\bar p^i_{k,\ell}$ with $p^i_{k,\ell}$ is small. With this respect, the discrepancy between the two densities  $p^i_{k,\ell}$ and $\bar p^i_{k,\ell}$ can be quantified
by computing the KLD
\begin{eqnarray}\label{eq:KLD}
D_{KL} (p^i_{k,\ell} \| \bar p^i_{k,\ell}) = \int p^i_{k,\ell} (x) \log \left ( {p^i_{k,\ell} (x) } / {\bar p^i_{k,\ell} (x)} \right ) \, dx \, ,
\end{eqnarray}
which, in Bayesian  statistics, represents precisely the information gain
achieved when moving from the old density $\bar p^i_{k,\ell} $ to the new one $p^i_{k,\ell}$.

Then, by considering the discrepancy measure (\ref{eq:KLD}), the following event-triggered transmission strategy is adopted
\begin{eqnarray}\label{eq:ts}
c^i_{k,\ell} = \left \{ 
\begin{array}{ll}  
0 & \mbox{if }  D_{KL} (p^i_{k,\ell} \| \bar p^i_{k,\ell})  \le \tau 
\\ 1 & \mbox{otherwise} 
\end{array} \right .
\end{eqnarray}
where the positive scalar $\tau$  can be seen as a design parameter that can be tuned so as to achieve a desired behavior in terms of transmission rate and performance.

Consider now the information fusion step. Clearly, when node $i$ receives the densities  $p^j_{k,\ell}$ from all its in-neighbors $j \in \mathcal N_i$, the fusion rule is the same as before. 
Instead, when $c^j_{k,\ell} = 0$ for  some neighbor $j$, then $p^j_{k,\ell}$  is not available and the fusion rule has to be modified. With this respect, note that in this case, thanks to the adopted event-triggered transmission
strategy (\ref{eq:ts}), node $i$ is still able to infer that the true $p^j_{k,\ell}$  is close  (in terms of KLD) to the reference density $\bar p^j_{k,\ell}$.
Then, a natural idea is to modify the information fusion step at node $i$ by replacing, for any $j \in \mathcal N_i$ such that
$c^j_{k,\ell} = 0$, the density $p^j_{k,\ell}$ with a suitable density $\tilde p^j_{k,\ell}$ computed from $\bar p^j_{k,\ell}$.
In fact, while in principle we could use directly $\bar p^j_{k,\ell}$ in the fusion step, it may be preferable to
modify it  so as to account for the additional uncertainty due to the discrepancy between $\bar p^j_{k,\ell}$ and $p^j_{k,\ell}$.
For example, this can be done by setting
\begin{eqnarray}\label{eq:tilde}
\tilde p^j_{k,\ell} (x) = \frac{ \left [ \bar p^j_{k,\ell} (x)  \right ]^{ \frac{1}{1+\delta}}   }{ \displaystyle{\int \left [ \bar p^j_{k,\ell} (x)  \right ]^{\frac{1}{1+\delta}} } dx }
\end{eqnarray}
with $ \delta \ge 0$, which corresponds to perform a flattening of the density  $\bar p^j_{k,\ell} $. To better understand this operation, we can observe that when $\bar p^j_{k,\ell}$ is a Gaussian with mean $\bar x^j_{k,\ell}$ and covariance
$\bar P^j_{k,\ell}$, then also $\tilde p^j_{k,\ell}$ will be a Gaussian with the same mean  $\tilde x^j_{k,\ell} = \bar x^j_{k,\ell} $ but increased covariance $\tilde P^j_{k,\ell} = (1+\delta) \bar P^j_{k,\ell} $, thus modelling the additional uncertainty.
In the following sections, we will show how, in the linear case, the scalar $\delta$ can be suitably tuned so as to ensure stability of the estimation error in all the network nodes.
Summing up, if we denote by ${\cal N}^i_{k,\ell}$ the set of in-neighbors of node $i$ for which $c^j_{k,\ell} = 1$,
each consensus step takes the form
\begin{eqnarray}
p_{k,\ell  + 1}^i\left( x \right) =   \frac{{  \left [ p^i_{k,\ell} (x) \right ]^{\pi_{i,i}} \displaystyle{\prod\limits_{j \in {{\cal N}^i_{k,\ell}} } {{{\left[ {p_{k,\ell }^j\left( x \right)} \right]}^{{\pi_{i,j}}}}} 
\displaystyle{\prod\limits_{j \in {\cal N}_i \setminus {{\cal N}^i_{k,\ell}} } {{{\left[ {\tilde p_{k,\ell }^j\left( x \right)} \right]}^{{\pi_{i,j}}}}}  }}}}{\displaystyle{\int { \left [ p^i_{k,\ell} (x) \right ]^{\pi_{i,i}} \prod\limits_{j \in {{\cal N}_{k,\ell}^i}} {{{\left[ {p_{k,\ell }^j\left( x \right)} \right]}^{{\pi_{i,j}}}}}   \displaystyle{\prod\limits_{j \in {\cal N}_i \setminus {{\cal N}^i_{k,\ell}} } {{{\left[ {\tilde p_{k,\ell }^j\left( x \right)} \right]}^{{\pi_{i,j}}}}}  }     dx} }}   
 \label{eq:ET-GCI}  
\end{eqnarray}

The above-described approach to  DSE with event-triggered communication gives rise to the algorithm of Table \ref{tab:2}.

\begin{table}[tb]
\caption{Algorithm 2 - Event-Triggered Distributed State Estimation with Consensus on Posteriors} \label{tab:2}
\hrulefill \\
At each time $k=0,1,\ldots$, for each node $i \in \mathcal N$:
\begin{enumerate}
\item {\bf Correction:} \\ {\bf If} $i \in \mathcal S$, collect the local measurement $y_k^i$ and update the local prior
$p^i_{k|k-1} $ via equation (\ref{eq:bayes:1})
to obtain the local posterior $p^{i}_{k|k}$; \\
otherwise, for any $i \in \mathcal N \setminus \mathcal S$, set
$p^{i}_{k|k} = p^i_{k|k-1} $;
\item {\bf Consensus:} \\
set $p^i_{k,0} = p^i_{k|k} $; \\
 {\bf For} $\ell  = 0, \ldots ,L-1$,  \\
 		\mbox{}	\quad \quad determine $c^i_{k,\ell}$ as in (\ref{eq:ts}); \\
 		\mbox{}	\quad \quad {\bf If}  $c^i_{k,\ell} = 1$ \\
		\mbox{}	\quad \quad \quad \quad transmit $p_{k,\ell}^i$ to the out-neighbors; \\
		\mbox{}	\quad \quad \quad \quad set $\bar p_{k,\ell+1}^i = p_{k,\ell}^i$; \\
		\mbox{}	\quad \quad {\bf Else} \\
		\mbox{}	\quad \quad \quad \quad set $\bar p_{k,\ell+1}^i = \bar p_{k,\ell}^i$; \\
		 \mbox{}	\quad \quad {\bf End if} \\
		\mbox{}	\quad \quad receive  $p_{k,\ell}^j$  from the in-neighbors $j \in {{\cal N}_i}$ for which $c^j_{k,\ell} = 1$; \\
		\mbox{}	\quad \quad {\bf For} all $j \in \mathcal N_i$ \\
		\mbox{}	\quad \quad \quad \quad {\bf If}  $c^j_{k,\ell} = 1$ \\
		\mbox{}	\quad \quad \quad \quad \quad \quad  set  $\bar p_{k,\ell+1}^j = p_{k,\ell}^j$; \\
		\mbox{}	\quad \quad \quad \quad  {\bf Else} \\
		\mbox{}	\quad \quad \quad \quad \quad \quad  set $\bar p_{k,\ell+1}^j = \bar p_{k,\ell}^j$; \\
		\mbox{}	\quad \quad \quad \quad \quad \quad  compute $\tilde p_{k,\ell}^j$ as in (\ref{eq:tilde}); \\
		\mbox{}	\quad \quad \quad \quad  {\bf End if} \\
		 \mbox{}	\quad \quad {\bf End for} \\		
		\mbox{}	\quad \quad	perform fusion using (\ref{eq:ET-GCI}); \\
		{\bf End for}
\item {\bf Prediction:} \\ compute the local prior $p^i_{k+1|k}$ from $p^{i}_{k,L}$ via equation (\ref{eq:bayes:2}); \\
compute the reference density $\bar p^i_{k+1,0}$ from $\bar p^{i}_{k,L}$ via equation (\ref{eq:bayes:2}); \\
compute the reference density $\bar p^j_{k+1,0}$ from $\bar p^j_{k,L}$ via equation (\ref{eq:bayes:2}) for any $j \in \mathcal N_i$.
\end{enumerate}
\hrulefill
\end{table}

\subsection{The linear-Gaussian case}

While in general implementation of Algorithm 2 requires some approximation, it turns out that all its steps admit a closed-form implementation
when the system is linear
\begin{eqnarray}
x_{k+1} & = & A \, x_k +  w_k
\label{LS1}
\\
y^i_{k} & = & \, C x_k + v^i_k \, , ~~~ i \in \mathcal{S} \, ,
\label{LS2}
\end{eqnarray}
and all the random variables, (i.e. the initial state, the process disturbance, and all the measurement noises) are normally distributed,
\begin{eqnarray*}
p_0 (x) &=& \mathcal G (x; \hat x_{0|-1},P_{0|-1}) \, , \\
p_w (w) &=& \mathcal G (w; 0,Q) \, ,\\
p_{v^i} (v^i) &=& \mathcal G (v^i; 0,R^i) \, , \quad i \in \mathcal S \, ,
\end{eqnarray*}
where: $\hat x_{0|-1}$ is a known vector and $P_{0|-1}, \, Q, \, R^i$, $i \in \mathcal S$, are known positive definite matrices;  $\mathcal G (\cdot ; \mu, \Sigma)$ denotes a Gaussian PDF with mean $\mu$ and covariance $\Sigma$.

In fact, as well known, thanks to the linear-Gaussian assumptions, the Bayesian filtering recursion
admits in this case a closed-form solution given by the Kalman filter recursion. This means that in the correction step,
given a Gaussian prior
\begin{eqnarray}
p^i_{k|k-1} (x) = \mathcal G (x; \hat x^i_{k|k-1},P^i_{k|k-1}) \, ,
\end{eqnarray}
the local posterior is again a Gaussian
\begin{eqnarray}
p^i_{k|k} (x) = \mathcal G (x; \hat x^i_{k|k},P^i_{k|k})
\end{eqnarray}
whose mean and covariance can be computed by means of the Kalman filter correction step (details are omitted since they are standard).

Further, also the consensus step preserves the Gaussian-form of the PDFs. To see this, it is convenient to consider, instead of mean and covariance, 
the information matrix
\begin{eqnarray}
\Omega^i_{k,\ell}  = (P^i_{k,\ell})^{-1} 
\end{eqnarray}
and information vector
\begin{eqnarray}
q^i_{k,\ell} = \Omega^i_{k,\ell} \hat x^i_{k,\ell} \, , 
\end{eqnarray}
which provide an alternative sufficient statistics for representing a Gaussian PDF. In fact, with some algebra, we can see that
the fusion step  (\ref{eq:ET-GCI}) preserves Gaussianity and can be written as a convex combination of the information pairs to be fused
\begin{eqnarray}
{q}_{k,\ell+1}^{i} &=& \pi_{i,i} \, {q}_{k,\ell}^i +  \sum_{j \in \mathcal{N}^i_{k,\ell}}~ \pi_{i,j} ~ {q}_{k,\ell}^j + \sum_{j \in \mathcal N_i \setminus \mathcal{N}^i_{k,\ell}}~ \pi_{i,j} ~ \tilde {q}_{k,\ell}^j   \label{CI:1} \\
{\Omega}_{k,\ell +1}^{i} &=&  \pi_{i,i} \, {\Omega}_{k,\ell}^i + \sum_{j \in \mathcal{N}^i_{k,\ell}}~ \pi_{i,j} ~{\Omega}_{k,\ell}^j  +  \sum_{j \in  \mathcal N_i \setminus \mathcal{N}^i_{k,\ell}}~ \pi_{i,j} ~ \tilde {\Omega}_{k,\ell}^j  \, . \label{CI:2}
\end{eqnarray}
where, for each $j \in \mathcal N_i \setminus \mathcal{N}^i_{k,\ell}$, the pair ($\tilde {q}_{k,\ell}^j , \tilde {\Omega}_{k,\ell}^j  $) is computed from the information pair ($\bar {q}_{k,\ell}^j , \bar {\Omega}_{k,\ell}^j  $) of the corresponding reference density $\bar p_{k,\ell}^j$ as
\begin{eqnarray}
\tilde {q}_{k,\ell}^j  &=& \frac{1}{1+\delta} \, \bar {q}_{k,\ell}^j  \label{eq:q:tilde} \\
\tilde {\Omega}_{k,\ell}^j &=& \frac{1}{1+\delta} \, \bar {\Omega}_{k,\ell}^j   \label{eq:O:tilde} 
\end{eqnarray}
Notice that equations (\ref{eq:q:tilde})-(\ref{eq:O:tilde}), which correspond to perform the flattening (\ref{eq:tilde}), basically amount to reducing the weights of the neighboring nodes that have not transmitted by a factor $1 +\delta$.

After consensus, the usual Kalman filter prediction step can be applied to the fused mean $\hat x^i_{k,L} = ( {\Omega}_{k,L}^i  )^{-1}  \, {q}_{k,L}^i $ and covariance $P^i_{k,L} =( {\Omega}_{k,L}^i  )^{-1} $ to get the predicted mean $ \hat x^i_{k+1|k}$ and covariance $P^i_{k+1|k}$.

Finally, notice that for Gaussian PDFs also the triggering condition (\ref{eq:ts}) can be evaluated in closed form since the KLD between $p^i_{k,\ell} $ and $\bar p^i_{k,\ell} $ can be written in terms of mean and inverse covariance as
\begin{eqnarray}
D_{KL} ( p^i_{k,\ell} || \bar p^i_{k,\ell} ) =  \frac{1}{2} \bigg \{ {\rm tr} [ \bar \Omega^{i}_{k,\ell} ( \Omega^i_{k,\ell} )^{-1} ] +  \| \hat x^i_{k,\ell} - \bar x_{k,\ell}^i \|^2_{\bar \Omega^i_{k,\ell}} 
 + \log \frac{\det \Omega^i_{k,\ell} }{\det \bar \Omega^i_{k,\ell}} - n \bigg \} \, ,
 \end{eqnarray}
where $n = \mbox{dim}(x)$.

\section{Stability analysis}

Focusing again on the linear-Gaussian case, we show now that, when the weight $\delta$ is sufficiently large, the proposed algorithm ensures stability of the estimation error in all network nodes under the minimal requirements of
network connectivity and collective observability. 

To this end, we first show that when the KLD between $p^i_{k,\ell}$ and $\bar p^i_{k,\ell}$ is small, i.e.
\begin{eqnarray}\label{eq:ts:2}
D_{KL}  (p^i_{k,\ell} || \bar p^i_{k,\ell} ) \le \tau
\end{eqnarray}
so that no transmission occurs, then also the true local estimate $\hat x^i_{k,\ell}$ and information matrix $\Omega^i_{k,\ell}$ are close to the estimate  $\bar x^i_{k,\ell}$ and information matrix $\bar \Omega^i_{k,\ell}$ provided 
by the reference density $\bar p^i_{k,\ell}$. \vspace{.3cm}

\begin{proposition} Let condition (\ref{eq:ts:2}) be satisfied. Then, there exist positive  scalars $\alpha^*(\tau)$, $\beta^* (\tau)$, and $\delta^* (\tau)$ such that
\begin{eqnarray}
 && \| \hat x^i_{k,\ell} - \bar x_{k,\ell}^i \|^2_{\Omega^i_{k,\ell}} \le \alpha^*(\tau)  \, , \label{eq:alpha} 
 \\ &&  \frac{1}{1+\beta^*(\tau)} \, \Omega^i_{k,\ell} \le \bar \Omega^i_{k,\ell} \le  (1 + \delta^*(\tau)) \, \Omega^i_{k,\ell} \, . \label{eq:beta}
\end{eqnarray}
\end{proposition} \vspace{.3cm}  
{\em Proof}: Notice first that, since $D_{KL} ( p^i_{k,\ell} || \bar p^i_{k,\ell} ) $ is always non-negative irrespectively of the values of $ \hat x^i_{k,\ell}$ and $\bar x^i_{k,\ell}$,
one has
\begin{eqnarray}
 \frac{1}{2} \bigg \{ {\rm tr} [ \bar \Omega^{i}_{k,\ell} ( \Omega^i_{k,\ell} )^{-1} ] + \log \frac{\det \Omega^i_{k,\ell} }{\det \bar \Omega^i_{k,\ell}} - n \bigg \} \ge 0
\end{eqnarray}
which implies
\begin{eqnarray}\label{eq:2tau}
 \| \hat x^i_{k,\ell} - \bar x_{k,\ell}^i \|^2_{\bar \Omega^i_{k,\ell}} \le 2 D_{KL} ( p^i_{k,\ell} || \bar p^i_{k,\ell}) \le 2 \tau \, .
\end{eqnarray}

Further, since  $\| \hat x^i_{k,\ell} - \bar x_{k,\ell}^i \|^2_{\bar \Omega^i_{k,\ell}} \ge 0$,  under condition (\ref{eq:ts:2}) we have
\begin{eqnarray}\label{eq:prop2:1}
{\rm tr} [ \bar \Omega^{i}_{k,\ell} ( \Omega^i_{k,\ell} )^{-1} ] + \log \frac{\det \Omega^i_{k,\ell} }{\det \bar \Omega^i_{k,\ell}} - n \le 2 D_{KL} (p^i_{k,\ell} || \bar p^i_{k,\ell} ) \le 2 \tau .
\end{eqnarray}
Notice now that, by exploiting the properties of matrix trace and determinant, the following identity can be derived
\begin{eqnarray} 
 {\rm tr} [ \bar \Omega^{i}_{k,\ell} ( \Omega^i_{k,\ell} )^{-1} ] + \log \frac{\det \Omega^i_{k,\ell} }{\det \bar \Omega^i_{k,\ell}}
= {\rm tr} [ ( \Omega^i_{k,\ell} )^{-1/2} \bar \Omega^{i}_{k,\ell} ( \Omega^i_{k,\ell} )^{-1/2} ]
- \log \det [( \Omega^i_{k,\ell} )^{-1/2} \bar \Omega^{i}_{k,\ell} ( \Omega^i_{k,\ell} )^{-1/2} ] \, .
\end{eqnarray}
Hence, inequality (\ref{eq:prop2:1}) can be rewritten as
\begin{eqnarray}\label{eq:prop2:2}
f \left [ ( \Omega^i_{k,\ell} )^{-1/2} \bar \Omega^{i}_{k,\ell} ( \Omega^i_{k,\ell} )^{-1/2} \right] \le 2 \tau
\end{eqnarray}
where $f(\cdot)$ is the matrix function
\begin{eqnarray}
f(X) = {\rm tr} (X) - \log \det (X) - n
\end{eqnarray}
defined over the cone of positive definite matrices. As it can be easily verified, the function $f(X)$ is convex and non-negative (it has a global minimum equal to $0$ in $X=I$).
Further,  $f(X)$ can be written in terms of the eigenvalues $\lambda_j$ of $X$ as follows
\begin{eqnarray}
f (X) = \sum_{j=1}^n ( \lambda_j -\log \lambda_j - 1 ) \, .
\end{eqnarray}
Since all the terms in the summation are nonnegative, $f(X) \le 2 \tau$ implies $\lambda_j -\log \lambda_j - 1 \le 2 \tau$ for any eigenvalue $\lambda_j$ of $X$.
Let us now denote by $\underline \lambda (\tau)$ and $\overline \lambda (\tau)$ the two solutions of the equation $\lambda -\log \lambda - 1 = 2 \tau$ where, for any $\tau >0$,
$0 < \underline \lambda (\tau) < 1 < \overline \lambda (\tau)$. It is an easy matter to check that  $\lambda_j -\log \lambda_j - 1 \le 2 \tau$ implies
$ \underline \lambda (\tau)  < \lambda_j < \overline \lambda (\tau)$ and, as a consequence, $f(X) \le 2 \tau$ implies $\underline \lambda (\tau) I \le X \le \overline \lambda (\tau) I$.
Hence, inequality (\ref{eq:prop2:2}) yields
\begin{eqnarray}
\underline \lambda (\tau) I  \le ( \Omega^i_{k,\ell} )^{-1/2} \bar \Omega^{i}_{k,\ell} ( \Omega^i_{k,\ell} )^{-1/2}  \le  \overline \lambda (\tau) I 
\end{eqnarray}
which can be written as in (\ref{eq:beta}) by letting $\delta^*(\tau) = \overline \lambda (\tau)  -1$ and $\beta^* (\tau) = 1/\underline \lambda (\tau) -1 $.

Finally, recalling (\ref{eq:2tau}), inequality (\ref{eq:alpha}) holds with $\alpha^* (\tau) = 2 \, \tau \, (1 + \beta^* (\tau))$. 
\qed
\vspace{.3cm}

A consequence of Proposition 1 is that, if we choose the scalar $\delta$ in the flattening step (\ref{eq:q:tilde})-(\ref{eq:O:tilde}) so that $\delta \ge \delta^* (\tau)$, then we have
\begin{eqnarray}\label{eq:delta:tau}
\tilde \Omega_{k,\ell}^i \le  \frac{1+\delta^* (\tau)}{1+\delta} \Omega_{k,\ell}^i \le \Omega_{k,\ell}^i  \, .
\end{eqnarray}
This condition is important because it ensures that  the information matrix after fusion is never larger than the one which would be obtained
in case all the nodes transmit, thus preventing the local filter from becoming too confident on the available information.
From the theoretical point of view, this property leads to the stability of the estimation error. To see this, 
let us consider the following assumptions.

\newcounter{totoro}
\vspace{0.3cm}
\begin{list}%
{\bf{A\arabic{totoro}.} }%
{\usecounter{totoro}%
\setcounter{totoro}{0}%
\setlength{\labelsep}{0pt} \setlength{\labelwidth}{25pt}
\setlength{\leftmargin}{25pt} }
\item The system matrix ${A}$ is invertible. \label{ass:inv}
\item The system is {\em collectively observable}, i.e. the pair $\, ({A},{C}) \,$ is observable where $ {C}
:= {\rm col} \left ({C}^i; \, i \in \mathcal S \right ) $. \label{ass:kalman}
\item The network is strongly connected, i.e., there exists a directed path between any pair of nodes $i,j \in \mathcal N$.
\label{ass:prim}
\end{list}\vspace{0.3cm}

Notice that these are the same assumptions under which stability of the Distributed Kalman filter with full transmission rate of Table \ref{tab:1} has been proved in 
\cite{battistelli2014kullback,battistelli2015consensus}. 
Notice also that assumption A\ref{ass:inv} is automatically satisfied in sampled-data systems
wherein the matrix ${A}$ is obtained by discretization of a continuous-time system matrix. 
Finally, let $\Pi$ denote the consensus matrix, whose elements are the consensus weights $\pi_{i,j} , \, i,j \in \mathcal N$ (in case $j \ne i$ does not belong to $\mathcal N_i$ we simply set $\pi_{i,j} = 0$); notice that assumption A\ref{ass:prim} ensures that $\Pi$ is primitive, i.e.
there exists an integer $\ell$ such that all the elements of ${\Pi}^\ell$ are strictly positive.  
Then, the following result can be stated.
\vspace{0.3cm}

\begin{theorem}
Consider the linear-Gaussian case and suppose that assumptions A1-A3 hold. 
Consider the estimates $\hat x^i_{k|k}$, $i \in \mathcal N$, generated by Algorithm 2 starting from positive definite  
information matrices $\Omega_{0|-1}^i$, $i \in \mathcal N$. 
Further, let  the  scalar $\delta$ in  (\ref{eq:q:tilde})-(\ref{eq:O:tilde}) be chosen so that $\delta \ge \delta^* (\tau)$. Then,
the estimation error is  uniformly bounded in mean square, i.e.
\begin{eqnarray}
 \limsup_{k \rightarrow \infty} \mathbb E \{ \| \hat x^i_{k|k} -x_k \|^2 \} < + \infty 
\end{eqnarray}
 in each network node $i \in \mathcal N$.
\end{theorem} \vspace{0.3cm}
{\em Proof:} The statement can be proved by following similar arguments as in the proof of Theorem 1 of \cite{battistelli2018distributed}. 
More specifically, in \cite{battistelli2018distributed} stability is proved by considering a triggering condition in which
no transmission occurs when both the current estimate $\hat x^i_{k,\ell} $ and inverse covariance $\Omega^i_{k,\ell}$ are close to the reference ones, i.e.
\begin{eqnarray}\label{eq:ts:3}
c^i_k = \left \{ 
\begin{array}{ll}  
0 & \mbox{if }   \| \hat x^i_{k|k} - \bar x_{k}^i \|^2_{\Omega^i_{k|k}} \le \alpha 
\mbox{ and }  \frac{1}{1+\beta} \, \Omega^i_{k|k} \le \bar \Omega^i_{k} \le  (1 + \delta) \, \Omega^i_{k|k}
\\ 1 & \mbox{otherwise} 
\end{array} \right .
\end{eqnarray}
where $\alpha$, $\beta$, and $\delta$ are positive scalars. 
While the transmission strategy (\ref{eq:ts:3}) is different from the one adopted here  (\ref{eq:ts}), Proposition 1 ensures that
the stability analysis of \cite{battistelli2018distributed} can be applied also in case of a transmission test defined directly in terms of KLD as in (\ref{eq:ts}), provided that
the scalars $\alpha$, $\beta$, and $\delta$ of \cite{battistelli2018distributed} are replaced by the scalars $\alpha^*(\tau)$, $\beta^*(\tau)$,  and $\delta^*(\tau)$ defined in Proposition 1.
\qed  \vspace{0.3cm}

Theorem 1 shows that the use of an event-triggered transmission strategy based on KLD does not destroy the stability properties of the DSE algorithm based on consensus on posteriors, 
while allowing for a reduction in the communication load.

\section{Performance evaluation}
In this section,
the performance of the proposed algorithm is checked via simulation experiments.
In our simulations,
the aim is to track a target moving inside a $5 \times 5$ $[km^2]$
surveillance area. 
The state of the target at time $k$ is defined as ${x_k} = [{{\xi _k}\;{{\dot \xi }_k}\;{\eta_k}\;{{\dot \eta}_k}}]{^\top} $,
where $[ {{\xi _k}\;{\eta_k}} ]{^\top}$ and $[ {{{\dot \xi }_k}\;{{\dot \eta }_k}} ]{^\top}$ 
denote respectively the target  position and velocity in Cartesian coordinates.
The target is supposed to move according to the constant-velocity kinematic model,
i.e. the matrix $A$ in (\ref{NLS1}) is given by
\begin{eqnarray}
A = \left[ {\begin{array}{*{20}{c}}
	1&T&0&0\\
	0&1&0&0\\
	0&0&1&T\\
	0&0&0&1
	\end{array}} \right],
\end{eqnarray}
where $T=1 \, [\mbox{s}]$ represents the sampling interval.
The covariance matrix $Q$ of the process noise is set to 
$Q = {\rm diag}([16 \, \mbox{m}^2, \, 1 \, \mbox{m}^2/\mbox{s}^2, \, 16 \, \mbox{m}^2, \, 1 \, \mbox{m}^2/\mbox{s}^2])$.

A network consisting of $100$ ($20$ sensor and $80$ communication) nodes is deployed over the surveillance area in order to track the moving target.
The location of the $i$-th sensor node is denoted by $[ {{\xi ^i}\;{\eta ^i}} ]^\top$. 
In our simulations,
two configurations of sensors are considered:
\begin{itemize}
	\item[-] {\bf Linear sensor case},
	where $10$ sensors measure the $\xi$-coordinate and other $10$ 
	the $\eta$-coordinate of  the target position.
	The measurement function (\ref{NLS2}) of each sensor node $i \in {\cal N}$ is, therefore, given by 
	\begin{eqnarray}
	y_k^i = \left\{ \begin{array}{l}
	{H_1}{x_k} + v_k^{i,\xi },\;\xi {\rm -coordinate} \\
	{H_2}{x_k} + v_k^{i,\eta },\;\eta {\rm -coordinate}
	\end{array} \right. 
	\end{eqnarray}
	where ${H_1} = \left[ {1,0,0,0} \right]$,
	${H_2} = \left[ {0,0,1,0} \right]$,
	and the variances of the measurement noises $v_k^{i,\xi }$, $v_k^{i,\eta }$
	are set to $R_k^{i,\xi } = R_k^{i,\eta }= 3 [{\rm m}^2]$.
	\item[-] {\bf Nonlinear sensor case},
	where $10$ sensors provide \textit{time-of-arrival} (TOA)
	and the other $10$ \textit{direction-of-arrival} (DOA) measurements.
	The measurement function (\ref{NLS2}) of each sensor node $i \in {\cal N}$ is, therefore, given by 
	\begin{eqnarray} \label{eq:MF}
	y_k^i = \left\{ \begin{array}{l}
		\sqrt {{{\left( {{\xi _k} - {\xi ^i}} \right)}^2} + {{\left( {{\eta _k} - {\eta ^i}} \right)}^2}} + v_k^{i,r} ,\;{\rm for\;TOA}                                       \\
		{\mathop{\rm atan}\nolimits} \left( {\frac{{{\eta _k} - {\eta ^i}}}{{{\xi _k} - {\xi ^i}}}} \right) + v_k^{i,\theta},\;\;\;\;\;\;\;\;\;\;\;\;\;\;\;\;\;\;\, {\rm for\;DOA}
	\end{array} \right.   
	\end{eqnarray}
	where the variance of the measurement noises $v_k^{i,r}$, $v_k^{i,\theta}$
	are set to $R_k^{i,r} = 9$ $[m^2]$ and $R _k^{i,\theta} = 0.01$ $[\deg^2]$, respectively.
\end{itemize}

\begin{figure}[tb]
	\centering
	\includegraphics[width=0.5\columnwidth]{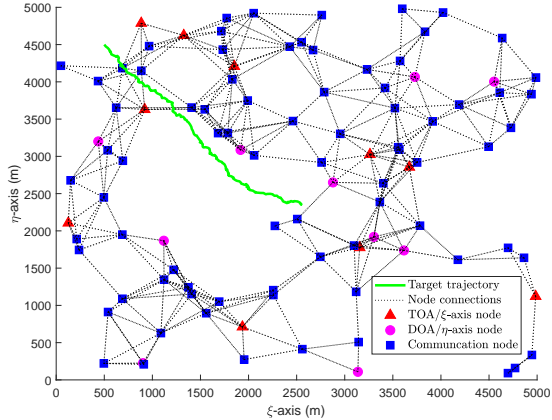}
	\caption{The simulated scenario.}
	\label{Fig:SCE}
\end{figure}

Details of the considered scenarios are illustrated in Figure \ref{Fig:SCE}.
Notice that for the nonlinear sensor case (\ref{eq:MF}),
the extended Kalman filter is adopted.
In order to better examine the performance of the proposed event-triggered strategy,
the simulation also involves other two transmission strategies:
a) randomly-triggered strategy, 
where each node randomly chooses the broadcasting time instants according to the preset transmission rate; and
b) periodically-triggered strategy,
where the message broadcasting time instants of each node are set in advance according to the transmission rate.
In the simulations, the transmission schedule is designed
to ensure that at least one node of the network will broadcast its message 
at each time instant under all selected transmission rates.
In all transmission strategies, the number of consensus steps is set to $L=1$.

\begin{figure*}[tb]
	\centering
	\begin{tabular}{cc}
	(a) & (b) \\
	\includegraphics[width=0.5\columnwidth]{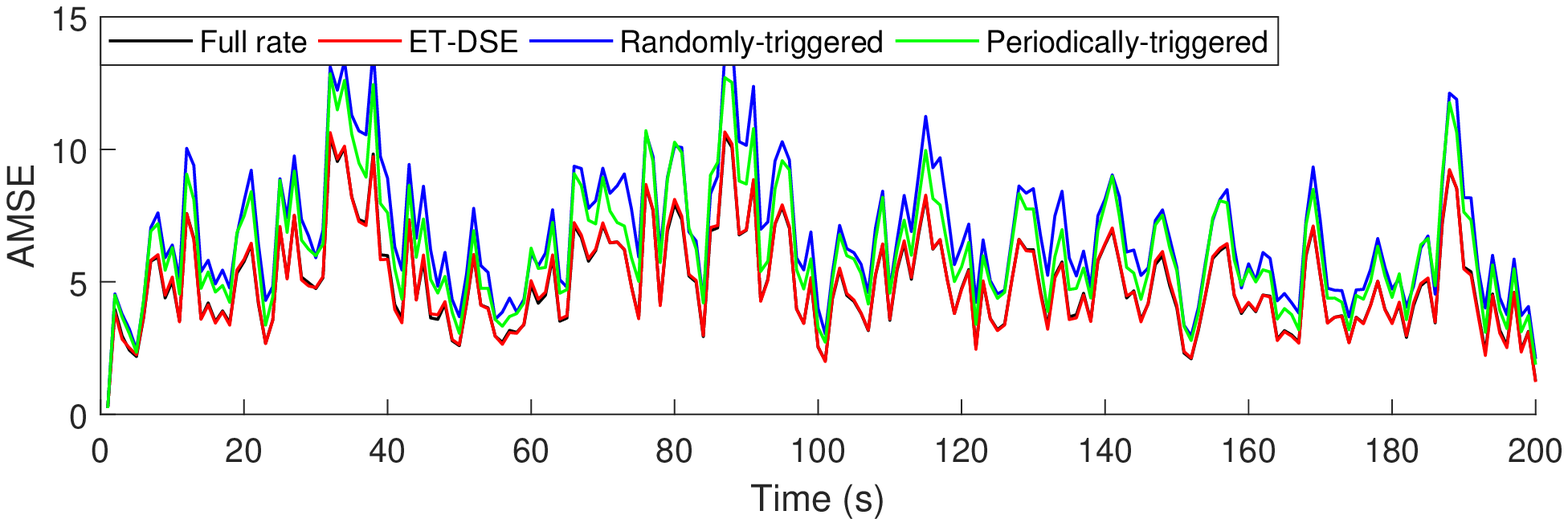} & \includegraphics[width=0.5\columnwidth]{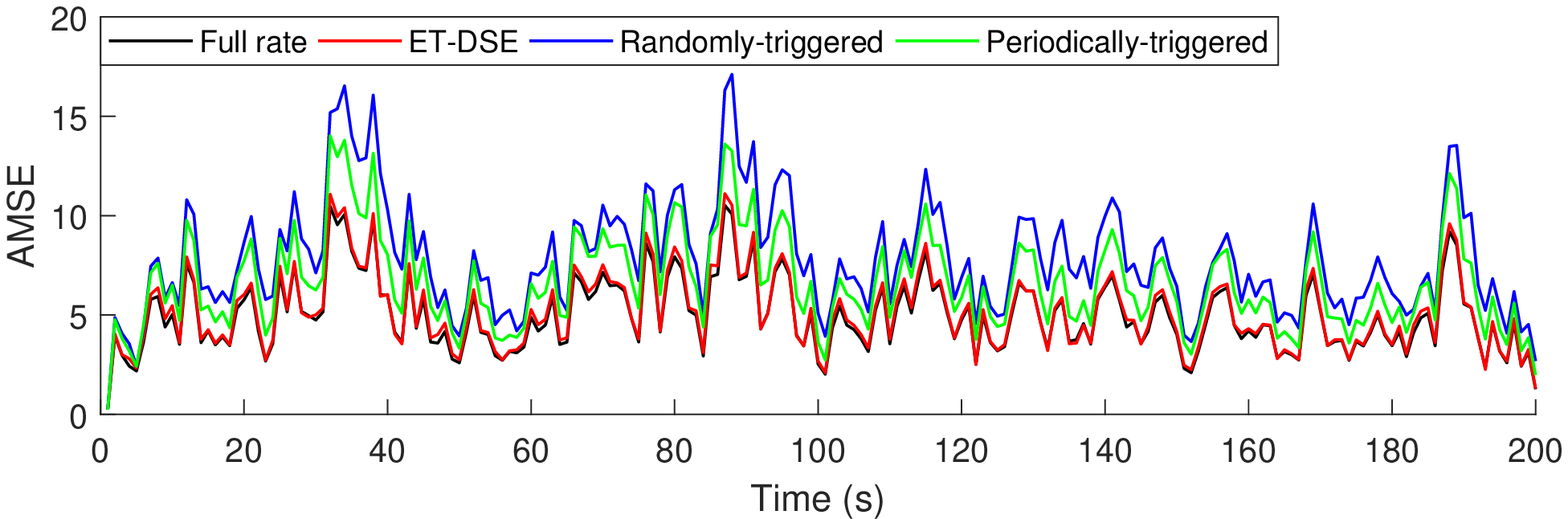} \\
	(c) & (d) \\
	\includegraphics[width=0.5\columnwidth]{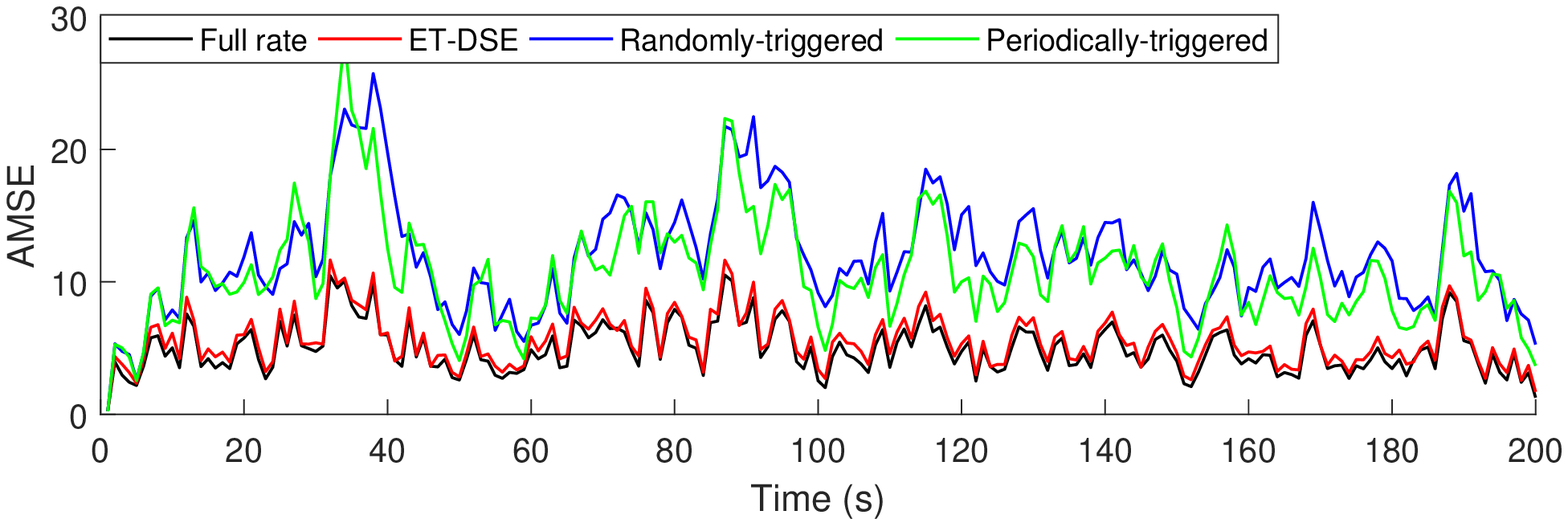} & \includegraphics[width=0.5\columnwidth]{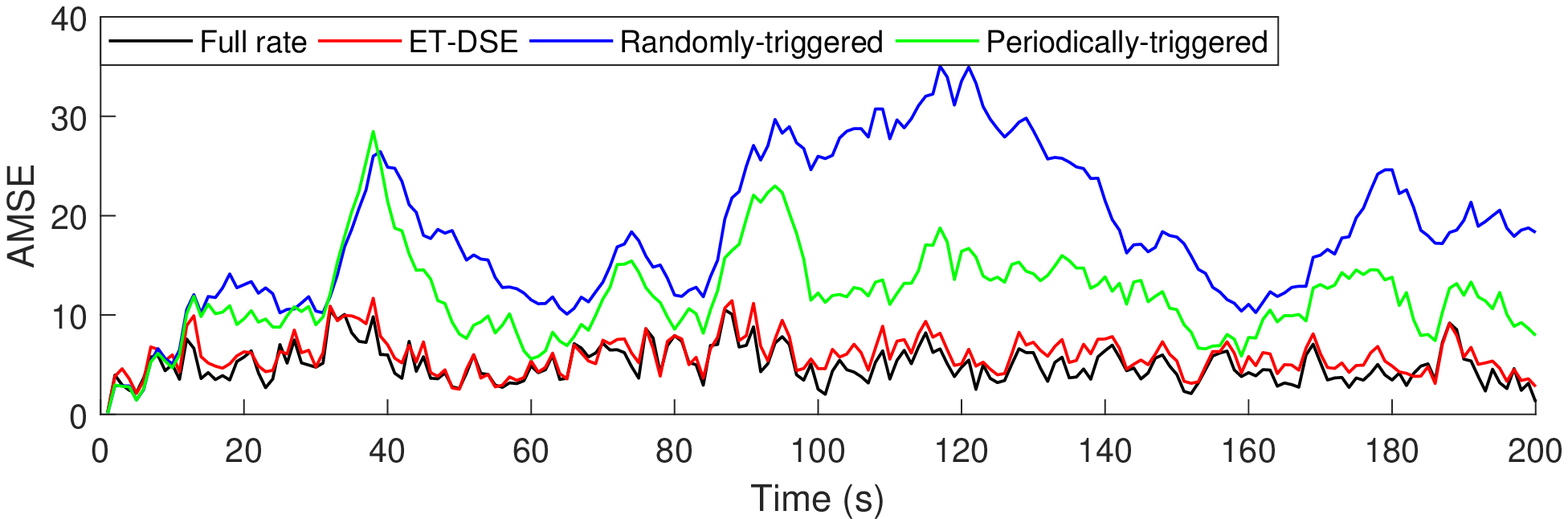} 
	\end{tabular}
	\caption{Linear sensor case -- performance evaluation under different communication rates: 70\% (a), 50\% (b), 30\% (c), and 10\% (d).}
	\label{Fig:L}
\end{figure*}

\begin{figure*}[tb]
	\centering
	\begin{tabular}{cc}
	(a) & (b) \\
	\includegraphics[width=0.5\columnwidth]{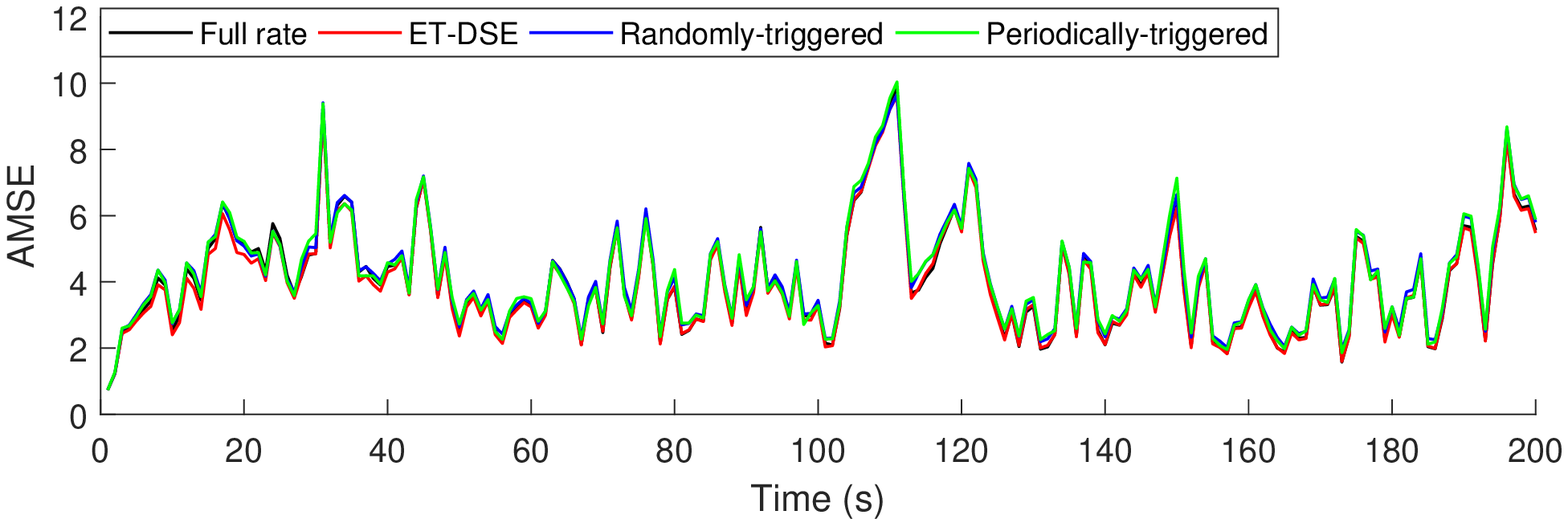} & \includegraphics[width=0.5\columnwidth]{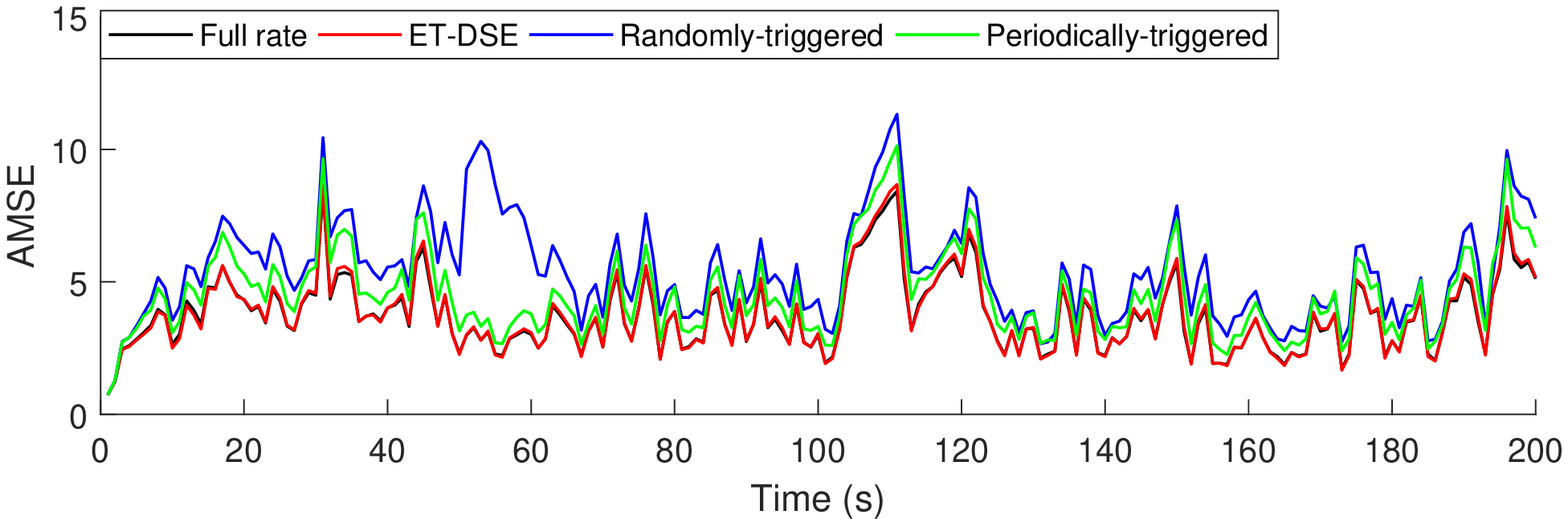} \\
	(c) & (d) \\
	\includegraphics[width=0.5\columnwidth]{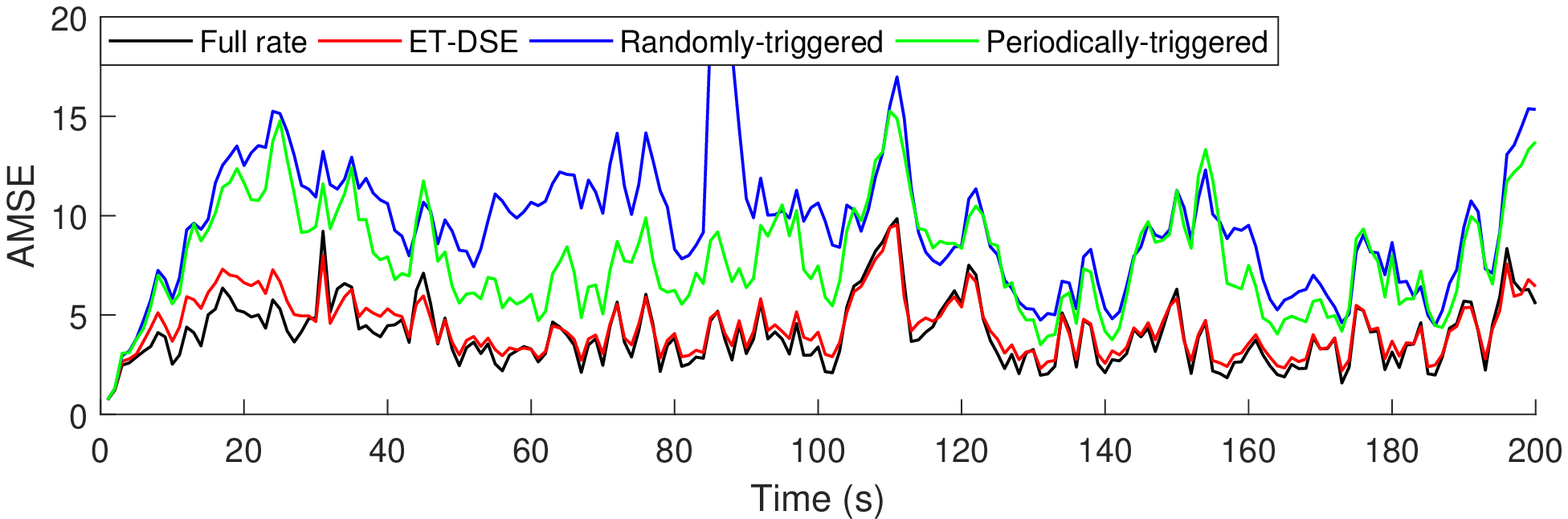} & \includegraphics[width=0.5\columnwidth]{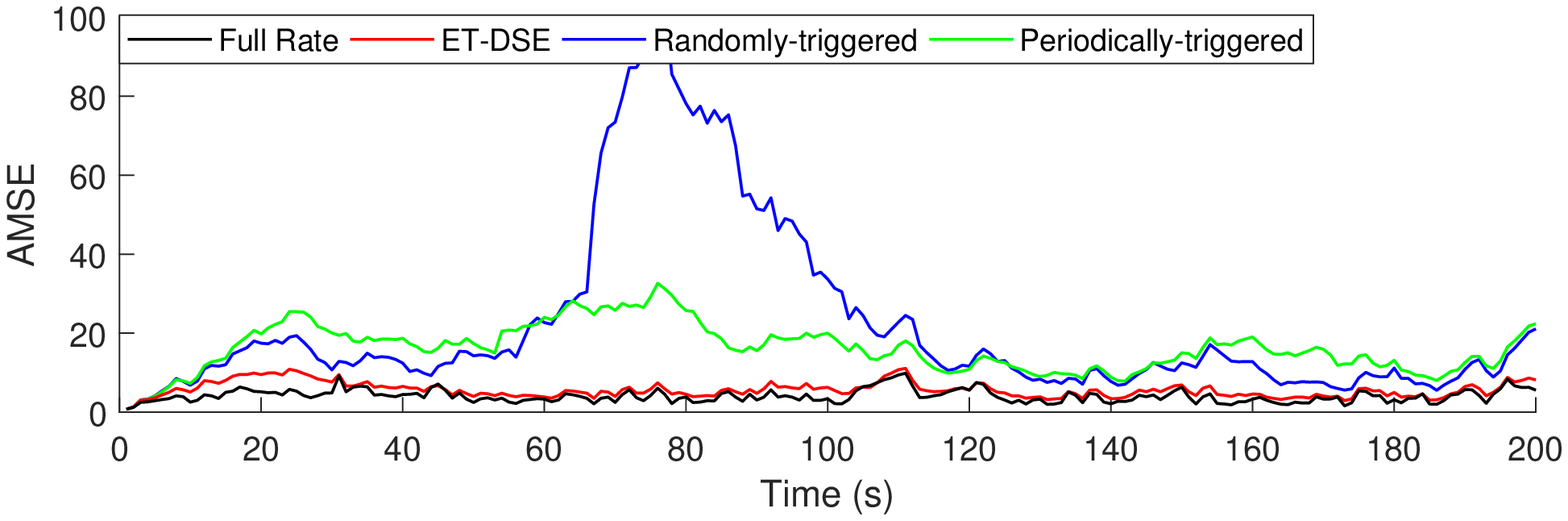} 
	\end{tabular}
	\caption{Nonlinear sensor case -- performance evaluation under different communication rates: 70\% (a), 50\% (b), 30\% (c), and 10\% (d).}
	\label{Fig:N}
\end{figure*}
%

As performance indicator, we employ the \textit{average mean square error} (AMSE) 
defined as follows:
\begin{eqnarray}
{{\cal E}_k} = \frac{1}{{\left| {\cal N} \right|}}\sum\limits_{i \in {\cal N}} {\left\| {\hat x_{k|k}^i - {x_k}} \right\|_2} ,
\end{eqnarray}
where ${\left\|  \cdot  \right\|_{\rm{2}}}$ denotes the Euclidean-norm.
In our simulations, 
$200$ independent Monte Carlo trials are carried out and the AMSE is further averaged with respect to the trials.
The performance achieved by the proposed event-triggered strategy at different transmission rates is illustrated in Fig. \ref{Fig:L}
for linear sensors
and in Fig. \ref{Fig:N}
for nonlinear sensors.
It can be concluded that
the proposed event-triggered strategy has better performance 
compared to the other two triggering strategies at the same transmission rate.
It can also be noticed that,
when the transmission rate increases, 
all triggering strategies perform close to the full-rate benchmark. 
In particular, the performance of the proposed ET-DSE algorithm
is always close to that of the full-rate one, 
even if the communication rate is extremely low (e.g. 30\%),
which means that the proposed event-triggered strategy 
can successfully balance estimation accuracy and energy consumption.

\section{Conclusion}
In this paper, an event-triggered consensus Bayes filter
is proposed in order to perform distributed state estimation by means of a sensor network, 
while reducing communication bandwidth and energy consumption
at each sensor node. 
The Kullback-Leibler divergence has been employed in the proposed event-triggered strategy 
in order to quantify the discrepancy between the local posterior distribution and the one predicted from the last transmission time. 
The effectiveness of the proposed approach has been demonstrated by computer simulations. 
Potential future research work will address the following issues:
(1) to develop a performance-predictable event-triggered strategy capable to
     adaptively choose the transmission triggering threshold so as to match a pre-specified transmission rate;
(2) to apply the proposed event-triggered strategy to distributed multitarget tracking and distributed multirobot \textit{simultaneous localization and mapping} (SLAM).

%

\end{document}